\documentclass[a4paper]{article}

\usepackage{graphics}
\usepackage{amsmath}
\usepackage{amssymb}
\usepackage{amscd}

\newcommand{\proof}{\textbf{Proof: }}
\newcommand{\cqfd}{\hfill \fbox{} \vskip 0.2cm}

\newcommand{\fig}[2]{\begin{figure}[!h] \centerline{\includegraphics{#1.eps}}
\caption{\label{fig_#1} #2} \end{figure}}

\begin{document}

\newtheorem{prop}{Proposition}
\newtheorem{theo}{Theorem}
\newtheorem{corol}{Corollary}
\newtheorem{lemme}{Lemma}
\newtheorem{definition}{Definition}
\newtheorem{notation}{Notation}
\newtheorem{rem}{Remark}


\begin{center}
{\huge \textbf{The lattice structure of\\ \vskip 0.2cm Chip Firing Games}}\\
\vskip 0.2cm
{\large{Matthieu Latapy and Ha Duong Phan\,\footnote{\textsc{liafa}, Universit\'e Paris 7, 2 
place Jussieu, 75005 Paris. (latapy,phan)@liafa.jussieu.fr}}}
\vskip 0.2cm
\end{center}

\noindent
\textbf{Abstract:} In this paper, we study a classical discrete dynamical
system, the Chip Firing Game, used as a model in physics,
economics and computer
science. We use order theory and show that the set of reachable states
(i.e. the configuration space)
of such a system started in any configuration is a lattice, which
implies strong structural properties.
The lattice structure of the configuration space of a dynamical
system is of great interest since it implies convergence (and more)
if the configuration space is finite. If it is infinite, this property
implies another kind of convergence: all the configurations reachable
from two given configurations are reachable from their infimum. In
other words, there is a unique first configuration which is reachable
from two given configurations.
Moreover, the
Chip Firing Game is a very general model, and we show how known
models can be encoded as Chip Firing Games, and how some results about
them can be deduced from this paper. Finally, we introduce a new model,
which is a generalization of the Chip Firing Game, and about which
many interesting questions arise.

\vskip 0.5cm

\section{Preliminaries}

A CFG (Chip Firing Game) \cite{BLS91} is defined over
a (directed) multigraph $G=(V,E)$,
called the \emph{support} or the \emph{base} of the game. A weight
$w(\nu)$ is associated with each vertex $\nu\ \in\ V$, which can
be regarded as the number of \emph{chips} stored at the \emph{site}
$\nu$. The CFG is then considered as a discrete dynamical system
with the following rule, called the \emph{firing rule}: a vertex containing
at least as many chips as its outgoing degree (its number of outgoing
edges) transfers one chip along each of its outgoing edges.
This rule can be applied in parallel (every vertex which
verifies the condition is fired at each step, see Figure~\ref{fig_ex_par})
or in sequential
(one vertex among the possible ones is fired, see Figure~\ref{fig_ex_seq}).
We can already observe a few points about the
example~: first, the total number of chips is constant, which is
obviously true for any CFG. Moreover, the example reaches a state
where no firing is possible. Such a state is called a \emph{fixed point}
of the CFG. Notice that there exists CFGs with no fixed point. In our
example, it seems that the third vertex (the one at the bottom) plays
an important role, as a collector.
Finally, if we fire one vertex at each step, like in
Figure~\ref{fig_ex_seq}, then
we may have to choose at some steps between two different vertices.
It is known from \cite{Eri93} that we will obtain a unique
fixed point (if any): the CFGs are strongly convergent games.
This means that either a CFG reaches
a fixed point, and then this point does not depend on the choices of
the vertices for firing, or the CFG has no fixed point.
Indeed, we have already seen that the two cases occur, and we will
explain more deeply in the following when each case occurs.

\fig{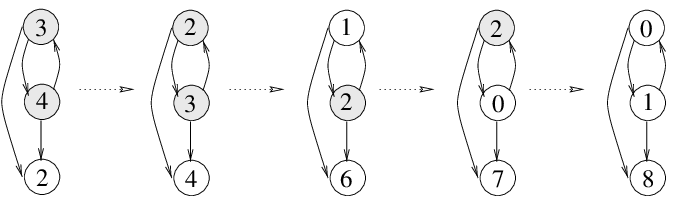}{An example of parallel CFG. The weight of each vertex is
indicated, and the shaded vertices are the ones
which can be fired.}

\fig{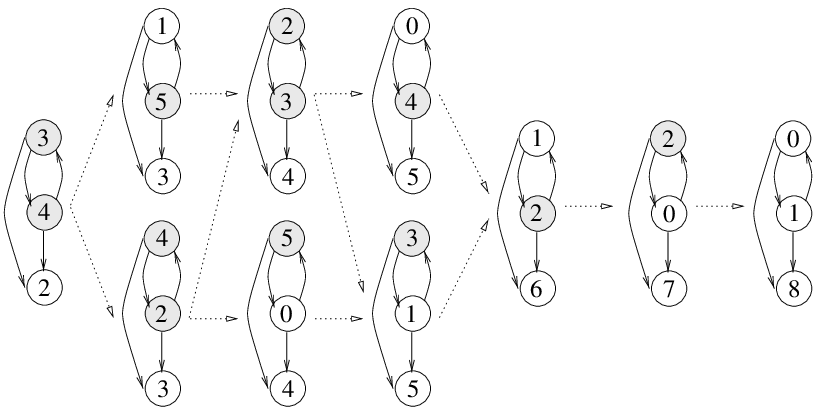}{An example of sequential CFG. The weight of each vertex is
indicated, and the shaded vertices are the ones
which can be fired.}

If we consider an arbitrary numbering of the vertices from $1$ to
$k$, we can obtain a description of a CFG with
the $k\times k$ matrix $M=(a_{ij})$ where $a_{ii}$ is
the number of outgoing edges from the vertex numbered $i$, and
$- a_{ij}$ is the number of edges from the vertex $j$
to the vertex $i$. The values of the other elements of
the matrix are $0$. This matrix
is known as the {\em laplacian matrix} of the support of the CFG \cite{BLS91}.
For example, for the CFG given in Figure~\ref{fig_ex_par}, $k=3$ and if we order
the vertices from the top to the bottom, we obtain~:
$$
M = \left( \begin{array}{ccc}
           2  & -1 & 0 \\
           -1 & 2  & 0 \\
           -1 & -1 & 0 
           \end{array} \right).
$$
The product of $M$ with a vector $v=\mbox{$^t$}(v_1,\dots,v_k)$
gives a vector $v'=\mbox{$^t$}(v'_1,\dots,v'_k)$
such that $v'_i$ is (the opposite of)
the variation of
the number of chips at vertex number $i$ if we fire $v_j$ times
vertex $j$, for all $1 \le j \le k$. Back to our example,
let us consider $v= \mbox{$^t$}(1,2,0)$. We have
$M\times v = \mbox{$^t$}(0,3,-3)$. This codes the variation
in the weight of each vertex
when we fire once the first vertex
and twice the second: the weight of the first vertex
is the same as before, while the weight of the second is
decreased by $3$ and the
weight of the last vertex is increased by $3$.
A \emph{sink} in a graph is a vertex such that there is no edge from it to
any other vertex. In the following, we will order the vertices of
a graph such that its sinks are the last vertices.
Therefore, if this graph has $m$ sinks, then the $m$ last columns of
its laplacian matrix contain only $0$.

CFGs are widely used in theoretical computer science, in physics
and in \mbox{economics}.
For example, CFGs model distributed behaviours
(such as dynamical distribution of
jobs over a network \cite{Hua93,DKTR95}), combinatorial objects (such as integer
partitions \cite{GK93,Bry73} and others \cite{CR98}). In physics, it is manly studied as a paradigm for so
called \emph{Self Organised Criticality}, an important area of research
\cite{BTW87,LMMP98,MN98}.
It was also proved in \cite{GM97} that (infinite) CFGs are Turing complete\,\footnote{This 
means that one can construct a CFG to simulate any Turing machine, and so there can not be any 
program which, given an infinite CFG and its initial configuration, says if the CFG will 
reach a stable configuration.},
which shows the potential complexity of their behaviours. However,
we will prove in the following that the set of possible configurations
for a CFG is
strongly structured. To acheive this, we will mainly use order theory.
A \emph{partially ordered set} (or \emph{poset}) is a set $P$ equipped with
a reflexive ($x \le x$), transitive ($x \le y$ and
$y \le z$ implies $x \le z$) and antisymetric ($x \le y$ and $y \le x$
implies $x=y$) binary relation $\le$.
A {\em lattice} is a poset such that two elements $a$ and $b$ admit a least
upper bound (called {\em supremum} of $a$ and $b$ and denoted by $\sup(a,b)$)
and a greatest lower bound (called {\em infimum} of $a$ and $b$ and denoted by
$\inf(a,b)$). The element $\sup(a,b)$ is the smallest element among the
elements greater than both $a$ and $b$. The element $\inf(a,b)$ is
defined dually. For more details, see for example \cite{DP90}.
The fact that the set of configurations of a dynamical system
naturally ordered is a lattice implies some
important properties, such as convergence. Moreover this convergence is
very strong in the following sense: for two configurations of the system,
there exists a unique first congiguration obtained from them, and every
configuration which can be obtained from both of them can be obtained
from this first one.

In the following, we will suppose (without loss of generality) that
the vertices of any graph are totally ordered, hence we will index
them with integers.
We suppose that this ordering is such that the sinks of
the CFG (i.e. the vertices with no outgoing edges) have the greatest indices.
Then, a configuration of a CFG is defined by a vector $c=(c_1,\dots,c_k)$
such that $c_i$ is the weight of the
$i$-th vertex.
A CFG defined over $G=(V,E)$ evolves from an initial configuration, and the set of all
configurations reachable from the initial one is called the
\emph{configuration space} of the CFG. This set is equipped with a
relation, called the \emph{successor relation}, induced by
the firing rule: $a \longrightarrow b$ if and only if the
configuration $b$ can be obtained from the configuration $a$
by firing one vertex of the CFG.
In other words, $a \longrightarrow b$ if there is one vertex $\nu$ in $V$
such that for all $v \in V, v \not= \nu$:
$b_v = a_v + | \lbrace (\nu,v) \in E \rbrace |$ and
$b_{\nu} = a_{\nu} - d^o_+(\nu) + | \lbrace (\nu,\nu) \in E \rbrace |$
The aim of this paper is mainly the
study of this configuration space and of this relation, which
gives deep insight about the behaviour of CFGs.

It will appear that a special class of CFGs play a central role for our
study. The supports of these CFGs do not have any \emph{closed component}~:
\begin{definition}[Closed Component]
A \emph{closed component} of a multigraph $G$ is
a nontrivial (more than one element) subset $S$ of the set of the
vertices of $G$ such that
\begin{itemize}
\item there exists a path from any element of $S$ to any other
      element of $S$ ($S$ is a nontrivial
      \emph{strongly connected component}), and
\item there is no outgoing edge from one element of $S$ to a vertex of $G$
      which is not in $S$ ($S$ is \emph{closed}).
\end{itemize}
\end{definition}
It is clear that the (support of the) CFG in Figure~\ref{fig_ex_par} has no
closed component, since its unique nontrivial strongly connected component is composed
of the two topmost vertices, and there is an edge from this component
to the third vertex, which is a sink.
Let us recall that when a chip arrives to such a sink, it
can never go out. Notice that a closed component behaves as 
a sink, since it also has this property. We will discuss more deeply
this analogy in the following. Continuing with our example,
if we delete this vertex then the graph is
reduced to a closed component, and we can notice that the obtained CFG has no fixed
point.

\section{CFG with no closed component}
\label{sec_noclosed}

In this section, we will show an important lemma which allows
the study of the special case where all the strongly
connected components of the support of the game have at least one
outgoing edge, i.e. an edge from one vertex in the component to
a vertex outside the component. This means that the support of
the game has no closed component. We show that in this case
the successor relation
of the CFGs induces an order over the configuration spaces.
Then, we extend the usual
definition (see for example \cite{Eri93,BLS91}) of
\emph{shot vectors} and we establish the
lattice structure of the configuration spaces of CFGs.

Let $G=(V,E)$ be the support multigraph of a CFG. Consider $G'$ the
quotient graph with respect to the (non closed) strongly
connected components:
the vertices of $G'$ are the strongly connected components of
$G$ and $C \longrightarrow D$ in $G'$ if and only if there
are vertices $a$ and $b$ of $G$ such that $a \in C$, $b \in D$,
and $a \longrightarrow b$ in $G$. The quotient graph $G'$ is
obviously a directed acyclic graph. 


We have the following lemma:

\begin{lemme}
\label{lem_cycle}
Let us consider a non closed strongly connected component $C$.
Starting from a configuration $a$ there is
no nonempty sequence of firings of vertices in $C$ such that
the configuration $a$ is reached again.
\end{lemme}
\proof
Suppose there is a nonempty sequence of firings of vertices in
$C$ which generates a cycle in the configuration space. Then,
denote by $v$ the first fired vertex. Likewise, denote
by $v'$ a vertex in $C$ such that there is a vertex $v''$
in a strongly connected component $D \not= C$ such that
$v' \longrightarrow v''$.
Since $v$ and $v'$ belong to the same strongly connected component $C$,
there is a path from $v$ to $v'$ in $G$:
$$
v = v_0 \longrightarrow v_1
        \longrightarrow v_2
        \longrightarrow
        \dots
        \longrightarrow v_k = v'
        \longrightarrow v''.
$$
After the firing of $v=v_0$, the weight of $v_1$ is increased
by at least one, hence there must be a firing of $v_1$ in order to complete
the cycle in the configuration space. Likewise, we have to fire
$v_2$, $\dots$, $v_k$. But the firing of $v_k=v'$ transfers one chip
from $C$ to $D$, which can obviously never come back by firing
only vertices in $C$. Hence we can
not complete the cycle in the configuration space, and we reach a
contradiction.
\cqfd

\noindent
This lemma implies that if we only fire vertices which are not in
a closed component, then we can not have any cycle of configurations.
Therefore, we deduce that if all the strongly connected components
which have no outgoing edge are trivial
(they contain only one vertex), i.e. there is no closed component, then
the configuration space of the whole CFG contains no
cycle, and so it is a poset:

\begin{theo}
The configuration space of a CFG with no closed component is partially ordered
by the reflexive and transitive closure of the successor relation.
\end{theo}

\noindent
Moreover, we know from \cite{Eri93} that the CFGs are strongly
convergent games. In other words, either a CFG does not converge at all,
either all the sequences of firings from one
configuration $a$ to another one $b$ have
equal length. Given such a sequence $p$, we denote by $\vert p \vert_i$ the
number of applications of the firing rule to the $i$-th vertex during
the sequence. Let us first prove the following result:

\begin{lemme}
\label{lem_top}
Given a CFG with no closed component, if, starting from the same
configuration, two sequences of firings $s$ and $t$ lead to the
same final configuration, then:
$$\vert s \vert_i = \vert t \vert_i \mbox{ for each vertex } i.$$
\end{lemme}
\proof 
Let $G=(V,E)$ be the support of this CFG and let $n= |V|$. 
Suppose that there exist two configurations $a$ and $b$
of this CFG such that there are two different sequences of firings
$s$ and $t$ from $a$ to $b$ which do not satisfy the condition
of the claim. Let us denote by $\sigma$ the vector
$(\vert s \vert_1,\vert s \vert_2,\dots,\vert s \vert_n)$ and
by $\tau$ the vector
$(\vert t \vert_1,\vert t \vert_2,\dots,\vert t \vert_n)$.
Let $M = (a_{ij})_{1 \leq i,j \leq n}$ be the laplacian matrix
of $G$. Recall that $- a_{ji}$ is equal to the number of edges
from $i$ to $j$ in $G$, and
$a_{ii} = \sum_{\stackrel{1\le j\le n}{j\not= i}} -a_{ji}$.
We know that $M\cdot \sigma = a-b \mbox{ and } M\cdot \tau = a-b$,
which implies $M\cdot (\sigma-\tau) = 0.$
Let us recall that the
sinks of $G$ are the last vertices $m+1, \ldots, n$ where $n-m$ is
the number of sinks of $G$ and so the matrix $M$ has the following form:
$$
M = \left( \begin{array}{ccccccc}
	a_{11}  & a_{12} & ... & a_{1m}& 0 & ... & 0 \\
	a_{21}  & a_{22} & ... & a_{2m}& 0 &  ... & 0 \\
 	...	& ... 	 & ... & ...   & 0  & ... & 0\\
 	a_{m1}  & a_{m2} & ... & a_{mm}& 0 & ... & 0 \\
 	a_{(m+1)1}  & a_{(m+1)2} & ... & a_{(m+1)m}& 0& ...  & 0 \\
	...	& ...	 & ...	& ...  & 0 & ...& 0 \\	
 	a_{n1}  & a_{n2} & ... & a_{nm}& 0 & ... & 0 	
           \end{array} \right).
$$
Remark also that for any sink $i$ (i.e. $m+1 \leq i \leq n$),
$\vert s\vert_i = \vert t\vert_i = 0$.
Let us denote by $\sigma'$ the vector
$(\vert s\vert_1, \ldots, \vert s\vert_m)$, by $\tau'$ the vector
$(\vert t\vert_1, \ldots, \vert t\vert_m)$, by $M'$
the matrix $(a_{ij})_{1 \leq i, j \leq m}$ and by $l_i$ the $i$-th
line of $M'$: $l_i = (a_{i1},\ldots, a_{im})$. We have $M\cdot (\sigma-\tau) =0$,
which implies that $M'(\sigma'-\tau')=0$, so $det(M')=0$
(since $\sigma \not= \tau$, and so $\sigma' \not= \tau'$) and
then there exists $m$ integers
$k_1, \ldots, k_m$ such that $k_1 l_1 + \dots + k_m l_m =(0, \ldots, 0)$.   

Let $i_1$ be an index such that $|k_{i_1}|$ is maximal between
$|k_1|, \ldots, |k_m|$. Notice that $i_1$ is not a sink since $i \leq m$. Without loss of generality, we can suppose
that $k_{i_1} > 0$. Let us consider the $i_1$-th column of $M'$; we have
the following inequality:
$$ k_{i_1} a_{i_1i_1} = k_{i_1} \sum_{\stackrel{j \not= i_1}{1 \le j \le n}}
- a_{ji_1} \ge k_{i_1} \sum_{\stackrel{j \not= i_1}{1 \le j \le m}}
- a_{ji_1} = \sum_{\stackrel{j \not= i_1}{1 \le j \le m}}
- k_{i_1} a_{ji_1} \ge \sum_{\stackrel{j \not= i_1}{1 \le j \le m}}
- k_j a_{ji_1}.$$
Moreover, by definition of $k_j$, we have $k_{i_1} a_{i_1i_1} = 
\sum_{\stackrel{j \not= i_1}{1 \le j \le m}} -k_j a_{ji_1}$, and so
the inequalities above actually are equalities. Therefore:
$$
\hspace*{-5cm}\left\{
\begin{array}{p{1cm}}
\begin{minipage}{1cm}$$\sum_{\stackrel{j \not= i_1}{1 \le j \le n}} -a_{ji_1} = 
\sum_{\stackrel{j \not= i_1}{1 \le j \le m}} - a_{ji_1}$$\end{minipage}\\
\begin{minipage}{1cm}$$k_j a_{ji_1} = k_{i_1} a_{ji_1} \mbox{ for all } 1 \le j \le 
m$$\end{minipage}
\end{array}
\right.
$$
which means that there is no edge from $i_1$ to any sink of $G$ and for
every $j$, $i_1 \rightarrow j$ in $G$ implies $k_j = k_{i_1}$.

Let us now consider a path from $i_1$ to a sink
$i_1 \rightarrow i_2 \ldots \rightarrow i_r \rightarrow p$
(notice that such a path always exists since $G$
has no closed component). The argument above says that
$k_{i_1} = k_{i_2} = \ldots = k_{i_r}$ and that there
is no edge from $i_1, i_2, \ldots, i_r$ to any  sink. But $p$ is
a sink and $i_r \rightarrow p$ is a edge in $G$, so we have a
contradiction. The proof is then complete.
\cqfd

\noindent
This lemma allows us to define the \emph{shot vector} $k(a,b)$ of two
configuration $a$ and $b$ if $b$ can be obtained from $a$ in a CFG: 
$k(a,b) = (k_1(a,b), \ldots, k_n(a,b))$ where $k_i(a,b)$ is the
number of firings of vertex $i$ to obtain $b$ from $a$.
Let us denote by $|k(a,b)|$ the sum $\sum_{i=1}^{i=n} k_i (a,b)$,
which is in fact the number of firings needed to obtain $b$
from $a$. 
If $a$ and $b$ are two configurations obtained from the same
configuration $O$, we order $k(O,a) \leq k(O,b)$ if
$\forall i$ $ k_i(O,a) \leq k_i(O,b)$. Moreover, if $a \geq b$,
it is clear that $k(O,b) = k(O,a)+k(a,b)$. Let us give here a
useful result about the shot vectors:
\begin{lemme}
\label{lem_3}
Let $a$ and $b$ be two configurations obtained from the same one
$O$ such that there exists an index $j$ such that
$k_{j}(O,a) \leq k_{j}(O,b)$ and
$ \forall j^{'} \neq j , k_{j^{'}}(O,a) \geq k_{j^{'}}(O,b)$.
If it is possible to fire the vertex $j$ of $b$,
then it is also possible to fire $a$ at the
same vertex.
\end{lemme}
\proof
Knowing that the necessary and sufficient condition to
fire the vertex $j$ of $b$ is $b_j \geq v_j$, where $v_j$
is the outgoing degree of $j$, let us consider $a_j$:
\begin{center}
\begin{tabular}{lcl}
$a_j$ & $=$    & $O_j - k_j(O,a) v_j + \sum_{i \rightarrow j \in G} k_i(O,a)$\\
      & $\geq$ & $O_j - k_j(O,b) v_j + \sum_{i \rightarrow j \in G} k_i(O,b)$\\
      & $=$    & $b_j$\\
      & $\geq$ & $v_j$.
\end{tabular}
\end{center}
So, it is possible to fire the vertex $j$ of $a$,
which proves the result.
\cqfd

\noindent
We can now characterize the order between all the configurations obtained
from the initial one $O$ in a CFG by comparing their shot vectors as follows:
\begin{theo}
\label{theo_2}
If $a$ and $b$ are two configurations obtained from the same
configuration $O$ of a CFG , then:
$$a \geq b \Longleftrightarrow k(O,a) \leq k(O,b).$$
\end{theo}
\proof
If $a \geq b$ then $k(O,b) = k(O,a)+k(a,b) \geq k(O,a)$. Let us
now assume that $k(O,a) \leq k(O,b)$  and consider two sequences
of firings, one from $O$ to $a$ and the other from $O$ to $b$:   
$$ O = c_0 \rightarrow c_{1} \rightarrow \ldots \rightarrow c_{r} \rightarrow a$$
$$ O = d_0 \rightarrow d_{1} \rightarrow \ldots \rightarrow d_{s} \rightarrow b.$$
We will construct step by step a sequence of firings
$ a \rightarrow e_{1} \rightarrow \ldots \rightarrow e_{t} \rightarrow b$,
showing that $a \geq b$. Knowing that
$(0,0,\ldots,0) =  k(O,O) \leq k(O,a) \leq k(O,b) $, there exists a first
configuration $d_i$ ($i \ge 1$) on the path from $O$ to $b$ such that
$k(O,d_i) \not\le k(O,a)$ and $k(O,d_{i-1} \le k(O,a)$.
Let $j$ be the vertex fired during the transition $d_{i-1} \rightarrow d_i$.
We have
$k_{j'}(O,d_{i-1}) = k_{j'}(O,d_{i})$ for all $j' \not= j$ and
$k_{j}(O,d_{i-1}) + 1 = k_{j}(O,d_{i})$.
Since $k(O,d_{i-1}) \le k(O,a)$, $k_{j'}(O,d_{i}) \le k_{j'}(O,a)$ for
all $j' \not= j$ and $k_j(O,d_i) \le k(O,a)+1$. But $k(O,d_i) \not= k(O,a)$,
and so $k_j(O,d_i) = k(O,a) + 1$.
Since $d_{i-1}$ and $a$
satisfy the conditions of Lemma~\ref{lem_3}, we can fire 
the vertex $j$ of $a$ to obtain a new configuration, denoted by
$e_{1}$, and we have $ k(O,d_{i}) \leq k(O,e_{1}) \leq k(O,b)$. By
iterating this procedure, we can define  $e_{2}, e_{3}, \ldots $
Since $|k(O,e_{l}) - k(O,a) | = l$ and
$k(O,a) \leq k(O,e_{l}) \leq k(O,b)$, after
$t = |k(O,b) - k(O,a) |$ steps we will have $ e_{t} \rightarrow b$.
A sequence of firings from $a$ to $b$ is then established.
\cqfd

\noindent
We can now state the main result of this paper:

\begin{theo}
\label{th_lat}
The set of all configurations obtained from the initial configuration $O$
of a CFG with no closed component, ordered with
the reflexive and transitive closure of the successor relation, is a
lattice. Moreover, the infimum of two elements $a$ and $b$ is defined as
follows:
let $k$ be a vector such that for all vertex $i$,
$k_i = max(k_i(O,a),k_i(O,b))$; then the configuration $c$ such
that $k(O,c) = k$ is the infimum of $a$ and $b$.
\end{theo}
\proof
Let us prove the given formula for the infimum. Then, since
a finite poset is a lattice if it contains a greatest element and
if it is closed for the infimum (see for example \cite{DP90}),
the fact that
$O$ is clearly the greatest element gives the result.

In order to prove that $c=\inf(a,b)$, we are going to show that
$a \geq c$ and $b \geq  c$. Since $c$ is clearly the greatest
configuration that can satisfy these properties, this will prove
the result. Let us assume that $k(O,a)$ and $k(O,b)$ are not
comparable (otherwise, $a$ and $b$ are comparable and the
result is obvious). Let us show that $a \geq  c$ (the proof
is similar for $b$). For that,  it is sufficient to find a
configuration $a'$ such that $a \rightarrow a'$ and
$k(O,a') \leq k(O,c)$. We are going to prove the existence
of such a configuration by using a sequence of firings
from $O$ to $b$. Let
$O \rightarrow d_{1} \rightarrow \ldots \rightarrow d_{s} \rightarrow b$
be such a sequence and let $i$ be the first index such that
$k(O,d_{i-1}) \leq k(O,a)$ and $k(O,d_i) \not\leq k(O,a)$.
Let us consider the vertex $j$ where the firing occurs
for $d_{i-1}$. We have $k_j(O,d_{i-1}) \leq k_j(O,a)$ and
$k_j(O,d_i) > k_j(O,a)$. Since $a$ and $d_i$ satisfy
the conditions of Lemma~\ref{lem_3}, we can fire the
vertex $j$ of $a$ to obtain a new partition $a'$. The
shot vector of $a'$ satisfies $\forall l \neq j$
$k_l(O,a') = k_l(O,a) \leq k_l(O,c)$ and
$k_j(O,a') = k_j(O,d_i) \leq k_j(O,b) \leq k_j(O,c)$.
By iterating this construction, we finally obtain a configuration
which verifies the wanted conditions.
\cqfd

\section{General CFG}

Our aim is now to study the structure of the configuration space
of any CFG, in particular the ones with closed component. We already noticed
that we do not obtain a lattice, and even not an order (see
Figure~\ref{fig_ex_cycle}(a)(b)). One natural way
to extend the results in the previous section is to consider the
quotient graph of the CFG with respect to the closed component:
each closed component $C$
is reduced to one vertex $c$ (which is obviously a sink) in the quotient
graph, and the initial CFG is reduced to the CFG over this graph
with the number of chips at $c$ being the total number of chips
in $C$. It is then obvious that the configuration space of the
quotient graph is a lattice, since this graph has no closed component
anymore. However, if we consider, for example, the graph in
Figure~\ref{fig_ex_cycle}(a), we can notice that, since the quotient graph
is reduced to one isolated vertex, we lose
a lot of information. Therefore, in this
section, we will give another natural extension of the notions 
presented in Section~\ref{sec_noclosed}. This extension will
also lead to lattice structures.

\fig{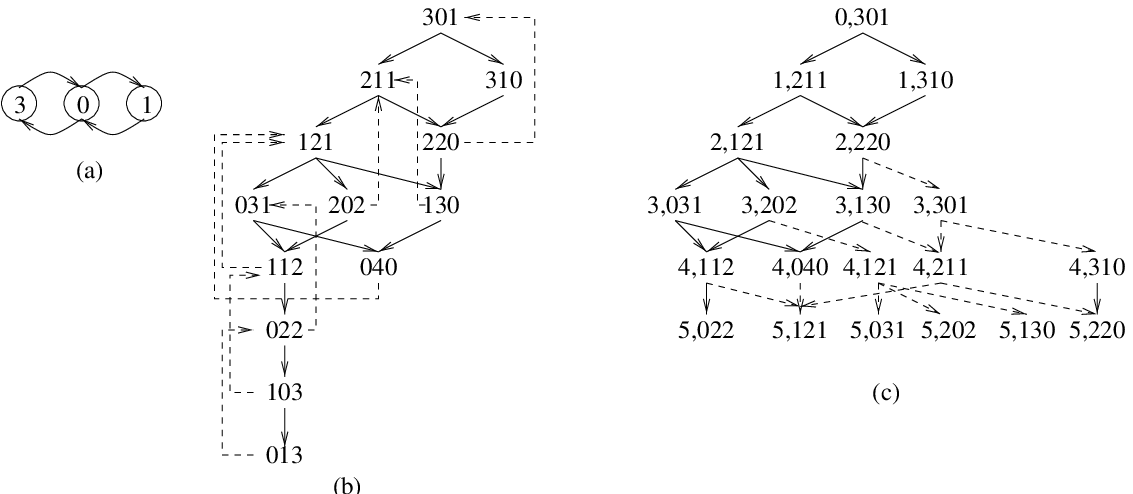}{An example of CFG with closed component:
(a) its support, (b) its configuration space  and (c) a
beginning of its extended configuration space.
The loops in the configuration space are represented by dotted lines.}

An \emph{extended} configuration of a CFG is a couple $(i,c)$ where
$c$ is a configuration, and $i$ is the total number
of firings used to obtain $c$ from the initial configuration. 
We naturally extend the notion of successor relation by saying that
$(i,a) \rightarrow (j,b)$ if and only if $j = i+1$ and $b$ can
be obtained from $a$ by \emph{one} application of the firing rule. See
for example Figure~\ref{fig_ex_cycle}(c). It is then obvious that the
extended configuration space of any CFG is an order.
Moreover, two sequences from
one extended configuration to another obviously have the same length.
We will now prove the
equivalent of Lemma~\ref{lem_top}, which is necessary to state
our result on CFG with closed component. Notice however that the proof
of this lemma is entirely different from the one of Lemma~\ref{lem_top},
and in fact it uses this lemma.

\begin{lemme}
Given a CFG, if, starting from the same extended
configuration, two sequences of firings $s$ and $t$ lead to the
same extended configuration, then:
$$\vert s \vert_i = \vert t \vert_i \mbox{ for each vertex } i.$$
\end{lemme}
\proof
Let $G=(V,E)$ be the support of the CFG we are studying.
Suppose that there exists two sequences $s$ and $t$ such that
there is $i$ such that $|s|_i \neq |t|_i$. Let us consider the
quotient graph $G'$ of $G$ with respect to its closed components
and the restricted CFG over $G'$. Since $G'$ has no closed component
then from Lemma \ref{lem_top}, ${i}$ cannot be a vertex of $G'$
(i.e. an equivalence class of cardinality $1$),
which implies that $i$ belongs to a closed component $C$
of $G$.

Without loss of generality, one can suppose that $|s|_i > |t|_i$
Let $S$ be the set of all the vertices $v$ of $C$ such that
$\vert s \vert_v > \vert t \vert_v$ ($S = C$ being clearly impossible).
Then, since $C$ is strongly connected, there must be at least one edge from
an element of $S$ to an element of $C \setminus S$.
Therefore, the set of vertices $S$ give more chips to the vertices
in $C \setminus S$ during $s$ than during $t$. Since at the end of
the two sequences the number of chips is the same,
this implies that the set $S$ must receive more chips during $s$ than
during $t$. Therefore, there is a vertex $v \not\in S$ such that $v$
gives more chips during $s$ than during $t$, which is a contradiction.
\cqfd

This lemma makes it possible to define the shot vector of
two extended configurations, and so we can prove the equivalent
of Lemma~\ref{lem_3} and Theorem~\ref{theo_2} in the context of
extended configuration spaces without changing the proofs.
We then obtain the following:

\begin{theo}
\label{th_lat2}
The set of all extended configurations obtained from the initial
extended configuration $(0,O)$ of a CFG with closed component, ordered with
the reflexive and transitive closure of the successor relation, is a
lattice. Moreover, the infimum of two elements $a$ and $b$ is defined as
follows:
let $k$ be a vector such that for all vertex $i$,
$k_i = max(k_i(O,a),k_i(O,b))$; then the extended configuration $(n,c)$
such that $k(O,(n,c)) = k$ and $n = \sum_{i\ge 1} k_i$ is the infimum
of $a$ and $b$.
\end{theo}
\proof
Similar to the proof of Theorem~\ref{th_lat} (in the case of infinte ordres, the existence of a greatest element and of an infimum for any subset of elements still implies the lattice structure). 
\cqfd

\section{Interesting special cases}

\label{sec_spm}

There exists many models which are in fact special Chip Firing Games.
The aim of this section is to show how some of these models belong to the class of CFG, and to show how known results can be deduced from the
results presented in this paper.

A first class of dynamical systems is used to modelize integer
partitions and granular systems in physics. They are composed of
a series of $k$ columns, each one containing a certain number of grains.
If the configuration is denoted by $a$, then we denote by $a_i$ the
number of grains at column $i$, and denote by $d_i(a)$ the integer
$a_i - a_{i+1}$ (with the assumption that $a_{k+1}=0$).
The system is then started with $n$ grains stacked in the
first column, and with no grain in the other columns.
Then, different evolution rules can be applied to the system, depending on
the studied model.
The most frequent one is SPM (Sand Pile Model), where
one grain in $a$ can be transferred from column $i$ to column $i+1$
if $d_i(a) \ge 2$ (see Figure \ref{fig_spm}(left)) \cite{GK93}.
We then obtain the
set $SPM(n)$ of reachable configurations from the initial one.
Two similar models have been developed as extensions of SPM:
$L(n,\theta)$ and $CFG(n,m)$. The model $L(n,\theta)$
simply consists in a variation of the threshold for the transfer of one
grain from one column to its right neighbour. Whereas it was $2$ in $SPM$,
it is $\theta$ in $L(n,\theta)$, where $\theta$ may be negative.
See Figure~\ref{fig_theta}(left). This
model was introduced and deeply studied in \cite{GMP00}.
The model $CFG(n,m)$ obeys another rule: let $a$ denote the
configuration of the the system, if $d_i(a) > m$ then $m$ grains
can fall from
column $i$ in such a way that each column $i+1$, $i+2$, $\dots$,
$i+m$ receives one grain \cite{GMP98b}. See Figure~\ref{fig_cfg}(left).

The two first models can be encoded as CFGs in the following way. Let $n$
be the number of grains in the system. Then, consider the
graph $G=(V,E)$ where $V = \lbrace 0,1,\dots,n \rbrace$ and
$E = \lbrace (i,i+1) \vert 1 \le i \le n-1 \rbrace \cup
     \lbrace (i,i-1) \vert n \ge i \ge 1 \rbrace$.
The vertex $i$ of this graph represents the column $i$ of the dynamical
system. Now, let $a$ be a configuration of the system, and suppose
we want to encode SPM. Then, we put $d_i(a)$ chips at vertex $i$ of the
CFG over G (with the assumption that $d_0(a)=0$), and we can verify that
the behaviour of the obtained CFG is equivalent to SPM (see
Figure~\ref{fig_spm}). This coding was first developed in
\cite{GK93}. Notice that
it is easy to reconstruct $a$ from the configuration of the CFG.
Likewise, $L(n,\theta)$ can be encoded
as a CFG in the same way as SPM, except that each vertex of the CFG
contains $d_i(a) - \theta + 2$ chips if $a$ is the corresponding
configuration of $L(n,\theta)$. See Figure~\ref{fig_theta}.
The support of the CFG that models the third model, namely $CFG(n,m)$,
is different: $V = \lbrace 0,1,\dots,n \rbrace$ and
each vertex $i$ has $m$ outgoing edges
$(i,i-1)$ and another outgoing edge $(i,i+m)$. See Figure~\ref{fig_cfg}.
A configuration $a$ of $CFG(n,m)$ is then equivalent to a configuration
of the Chip Firing Game where each vertex $i$ contains $d_i(a)$ chips.

\fig{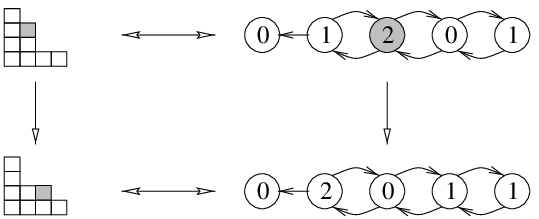}{Coding of the Sand Pile Model with a Chip Firing Game.}

\fig{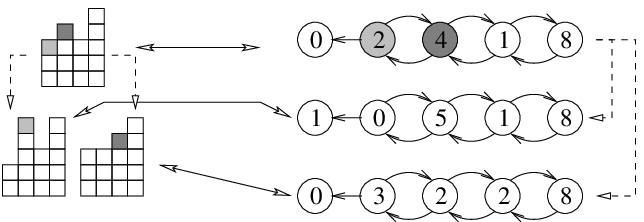}{Coding of $L(n,\theta)$ with a Chip Firing Game
 when $\theta = -1$.}

\fig{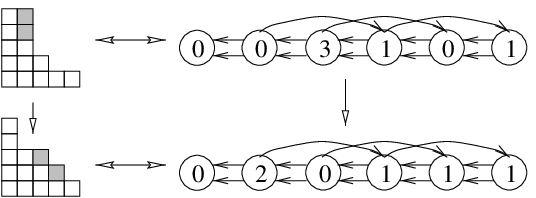}{Coding of $CFG(n,m)$ with a Chip Firing Game when $m = 2$.}

Notice that the CFGs used to encode SPM, $L(n,\theta)$ and $CFG(n,m)$
contains no closed component.
Therefore, we obtain the following result, previously known from
\cite{GK93}, \cite{GMP98b} and \cite{GMP00},
as a corollary of Theorem~\ref{th_lat}:
\begin{corol}[\cite{GK93,GMP98b,GMP00}]\ \\
The sets $SPM(n)$, $L(n,\theta)$ and $CFG(n,m)$ equipped with
the reflexive and transitive closure of the successor relation
are lattices, and these dynamical systems converge to a unique fixed point
independant of the sequence of applications of the rule used. Moreover,
all the paths from one configuration to another have the same length
and involve the same applications of the rule.
\end{corol}

A similar model, the Game of Cards, was
introduced in \cite{DKTR95} to study a
distributed algorithm. The game is very
simple: it is composed of $k$ players disposed on a ring, and each
player can give a card to its right neighbour if he/she has more cards than
him/her. The corresponding CFG is a ring of $k$ vertices: the $i$-th vertex
has an outgoing edge to vertex $i+1$ modulus $k$ and another one
to $i-1$ modulus $k$. Then, a configuration $a$ of the game is
encoded by a configuration of the CFG where vertex $i$ contains as 
many chips as the difference between the number of cards of player $i$
and the number of cards of its right neighbour plus $1$. Notice that
this codage is quite different from the previous ones, since the graph of the 
obtained CFG is a cycle and then it is itself  a closed component. Therefore, we apply Theorem~\ref{th_lat2}
and we obtain:
\begin{corol}[\cite{GMP98a,DKTR95}]
The set of extended configurations of the Game of Cards is a lattice.
Moreover, each sequence of plays between two states have the same
length and involve the same players.
\end{corol}

Our last example is the Abelian Sandpile Model indroduced in
physics to represent typical behaviours in granular
systems and self-organised criticality \cite{DM90, DRSV95}.
The system is defined over an undirected graph with
one distinguished vertex, called the {\em sink}, or the {\em exterior}.
Each vertex contains a number of grains, and it can give
one grain to each of its neighbours if it has more grains than
its degree. This rule, called
the \emph{toppling rule}, is
applied to every vertex except the sink. See Figure~\ref{fig_cori}.

\fig{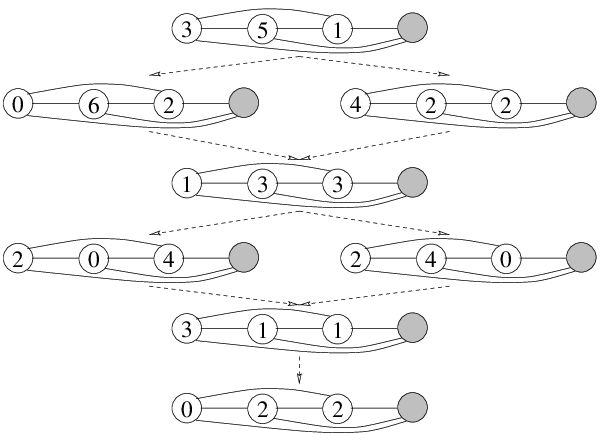}{Behaviour of an Abelian Sandpile Model. The sink is the
shaded vertex. This is equivalent to a CFG, hence the set of
configurations is naturally ordered as a lattice.}

The underlying graphs
of the original model were $d$-dimensionnal regular
finite grids with the sink representing the exterior. The model
have then been extended to any graph with a sink and nice algebraic
results have been developed \cite{CR98, Mar92}. However, the fact
that the configuration space of such systems is always a lattice
is a new result induced by Theorem \ref{th_lat}. Indeed, let us consider
an Abelian Sandpile Model over a graph $G$ and let us construct
a CFG with support $G'$ such that the two systems are equivalent.
Let $G'$ defined by:
\begin{itemize}
\item If there is an undirected edge $\lbrace i,j \rbrace$
in $G$ then $G'$ contains the directed edges $(i,j)$ and $(j,i)$ in $G'$,
except if either $i$ or $j$ is the sink.
\item If there is an undirected edge $\lbrace i,j \rbrace$
in $G$ such that $j$ is the sink, then $G'$ contains the directed
edge $(i,j)$.
\end{itemize}
Now it is obvious that this CFG is equivalent to the Abelian Sandpile
Model when a grain is identified with a chip.
Therefore, we obtain the convergence and the fact that
all the paths from one configuration to another have
the same length and involve the same applications of the rule,
which is proved in \cite{DM90, DRSV95, CR98, Mar92}, and one more
result:
\begin{corol}
Given an initial configuration, the set of reachable configuration
of an Abelian Sandpile Model is a lattice.
\end{corol}

For related models used in economics, and for which many results can
be obtained from the ones presented here, we send the reader to 
\cite{Heu99,Big97,Big99}. The models developed in these papers are very
close to our Chip Firing Game, and the questions they study
concern convergence, length of paths, and others.

\section{A Note about Parallel versus Sequential CFG}

The results presented until now deal with the sequential Chip Firing
Game: at each step, we apply the firing rule to only \emph{one}
vertex among the possible ones. However, as we already noticed in
the introduction, the parallel case, where we apply the firing
rule to each vertex at each step, is of great interest: it represents
many physical phenomenons and is a model for distributed algorithms.
We will now introduce two new models which add a certain amount of
parallelism to CFGs. These two generalizations seem relevant since
many practical problems can be modelized this way.

The first step is the \emph{semi-parallel} Chip Firing Game,
denoted by $CFG(k)$, where
at each step we apply the firing rule to at most $k$ vertices among
the possible ones. If $k=1$, we simply obtain the sequential model.
However, if $k \ge 2$, this model creates transitivity edges in the
configuration space. Indeed, if for example two firings are possible
from a configuration $c$, then $c$ has three successors: $c \rightarrow c'$,
$c \rightarrow c''$ and $c \rightarrow c'''$ where $c'$ is obtained from
$c$ by firing one of the two possible vertices, say $\nu'$,
$c''$ is obtained
from $c$ by firing the other possible one, $\nu''$, and $c'''$ is obtained
from $c$ by firing simultaneously the two vertices $\nu'$ and $\nu''$.
Notice that it is possible to fire the vertex $\nu''$ from $c'$, and
this leads to $c'''$. Likewise, it is possible to fire $\nu'$ from $c''$
in order to obtain $c'''$. Therefore, the edge $c \rightarrow c'''$ is simply
a transitivity edge, and one can verify that the same phenomenon appears
if there are more than two vertices which can be fired, and for
any $k\ge 2$.
See Figure~\ref{fig_semipar} for an example.
This immediately implies that the semi-parallel $CFG(k)$ lead to the
same final state (if any), independently of $k$.

\fig{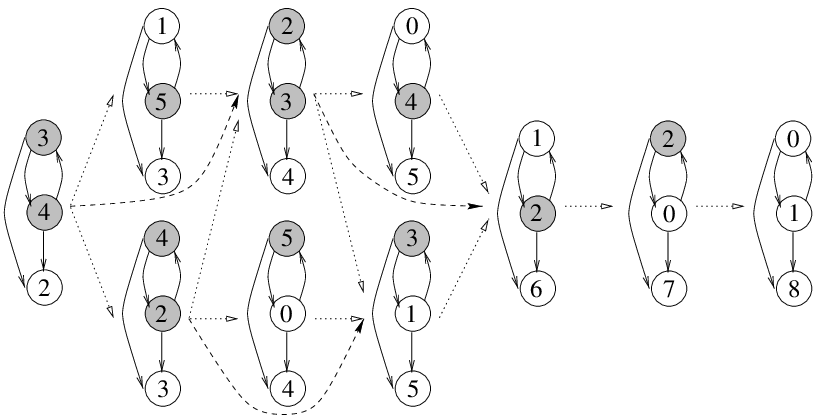}{The configuration space of a semi-parallel CFG with 
$k=2$.}

Now, a natural variation is the \emph{maximal semi-parallel}
Chip Firing Game, where we fire as much vertices as possible but no more
than $k$. This model is exactly the parallel Chip Firing Game
if $k$ is equal to the number of vertices of the support. On the other
hand,
we have the sequential Chip Firing Game when $k=1$. Moreover, it is clear
that the configuration space of a maximal semi-parallel CFG is obtained
from the configuration space of the corresponding semi-parallel CFG by
deleting all the configurations $c_1$, $c_2$, $\dots$, $c_l$ such that
the transitivity edge $c_0 \rightarrow c_{l+1}$ exists, with
$c_0 \rightarrow c_1$ and $c_l \rightarrow c_{l+1}$, and then taking
the configurations which are reachable from the initial one
(see Figure~\ref{fig_max_semipar}).
Then, it is clear that the configuration space of a maximal semi-parallel
CFG is a subset of the configuration space of its corresponding sequential
CFG, and that both have the same fixed point.
Therefore, both sequential, semi-parallel,
maximal semi-parallel and parallel CFG have the same fixed point.

\fig{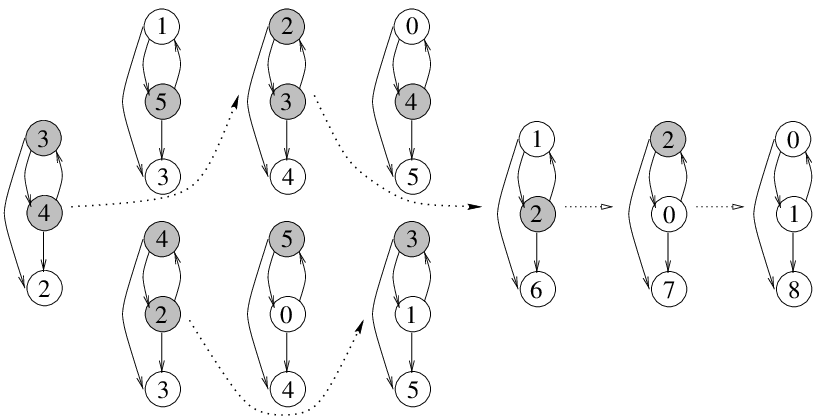}{Construction of the configuration space of a maximal
semi-parallel CFG with $k=2$. After deleting the edges for
which there exists shortcuts, we simply take the configurations
reachable from the initial one. Here, we finally
obtain the same configuration space as in the parallel case
(see Figure~\ref{fig_ex_par}).}


It seems now natural to study how the lattice properties
evolves when we consider maximal semi-parallel CFGs instead
of sequential ones.
It should be natural to expect that the configuration space
of such a game is a sub-lattice of the
corresponding sequential CFG. Actually, as shown in
Figure~\ref{fig_contrex}, this is not true
since the configuration space of a
maximal semi-parallel CFG is not (in this case) a sub-order of
its corresponding sequential game. We can say even more from this simple
example: in a maximal semi-parallel game, there exist paths of
different lengths from
one configuration to another.
This is a very
important remark, which shows that
some maximal semi-parallel can not be simulated by any
sequential CFG. Moreover, if each edge consists in
an elementary operation in a (parallel) computer, then
we can manage the calculus necessary to
go from one configuration to another in an efficient way.
Finally, we can see on Figure~\ref{fig_contrex2} that the configuration
space of a maximal semi-parallel CFG is not always a lattice.

\fig{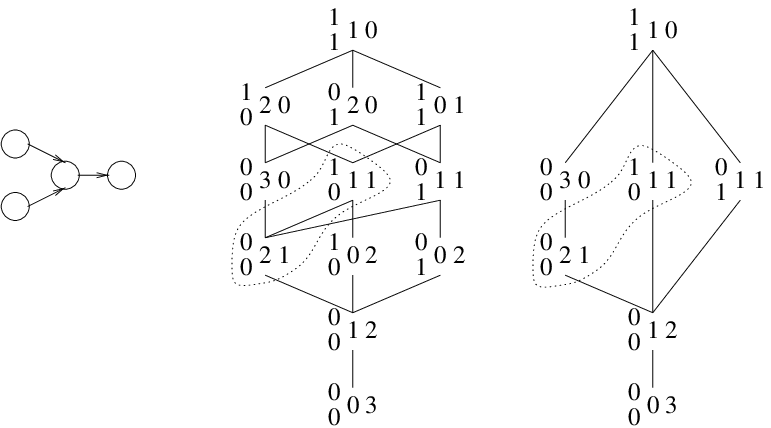}{From left to right: the support of the considered CFG;
the configuration space of the sequential CFG started in a particular
configuration; the configuration space of the corresponding
maximal semi-parallel CFG(k) with $k=2$. The two outlined configurations
show that the order is not preserved.}

\fig{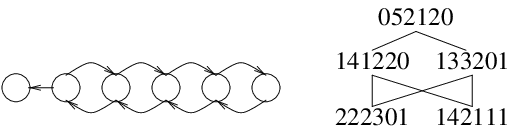}{A maximal semi-parallel CFG for which the configuration
space is not a lattice when $k=2$. Notice that the support is very simple:
it is nothing but the equivalent of SPM (see Section~\ref{sec_spm}).}

\section{Perspectives}

This paper shows that the structure of the configuration space of
any Chip Firing Game can be viewed as a lattice. This explains
the strong relation between many dynamical systems and lattices 
already noticed in previous papers. Moreover, the model of Chip
Firing Games is very general, which leads to the possibility of
proving the lattice structure of other models by coding them as
special Chip Firing Games.

This also raise to another kind of questions: since the Chip Firing
Games are very general, many lattices can be viewed as the
configuration space of a Chip Firing Game. Lemma~\ref{lem_top} also
shows that two sequences of firings from a configuration to another
one have the same length, which means that the obtained lattices
are \emph{ranked}\footnote{A lattice is \emph{ranked} if, in the
graph of the covering 
relation of its order, two directed paths from one element to another have
the same length \cite{DP90}.}. Moreover, if $a \rightarrow b$ and
$a \rightarrow c$ in the configuration space of a CFG, then
there exists $d$ such that $b \rightarrow d$ and $c \rightarrow d$.
Therefore, there exists some lattices
which are not isomorphic to any configuration space of a CFG.
However, we do not know if the class of
the lattices which verify these two properties corresponds exactly to
the class of these lattices isomorphic to the configuration
space of a CFG. If it is not the case, it should be very
interesting to look for a characterization of these lattices.
Likewise, the special case of \emph{distributive}\footnote{A lattice
is \emph{distributive} if for all $a$, $b$ and $c$:
$\sup(\inf(a,b),\inf(a,c)) = \inf(a,\sup(b,c))$ and 
$\inf(\sup(a,b),\sup(a,c)) = \sup(a,\inf(b,c))$.
For more details, see
\cite{DP90}.}
lattices is very interesting. Is any distributive lattice isomorphic
to the configuration space of a CFG ? For example, we show in
Figure~\ref{fig_ideal} the CFG equivalent to the distributive
lattice obtained from the empty partition by iteration of the
following rule: one can add one grain to a column of a partition if one
obtains this way another partition. This lattice can also be thought
as the lattice of the ideals of the product of two chains, which
is known to be distributive \cite{DP90}.

\fig{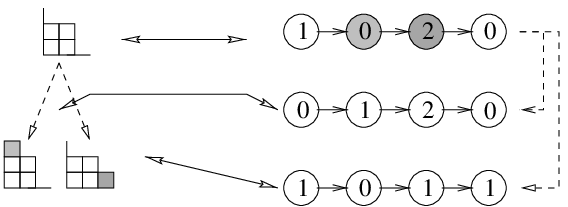}{The coding of the partitions with addition of one grain,
i.e. the ideals of the product of two chains (here,
$\lbrace {1,2,3} \rbrace \times \lbrace {1,2,3} \rbrace$), with a
Chip Firing Game.}


Finally, an interesting open problem was pointed out by Moore
and Nilsson \cite{MN98}. They studied the special case of Abelian Sandpile
Model on $d$-dimensional grids. If $d=1$, this is nothing but
$SPM$. If $d\ge 3$, they proved that the problem of calculating
the final state of the system started from an arbitrary configuration
is P-complete. However, the case $d=2$ is still open, and it
is a challenge to determine the complexity of the prediction of
the final state in this case. We show here that the configuration
space in each of these cases is a lattice. It should be worth
to study more deeply the structure of the obtained lattice
in the case $d=2$, which would lead to insight in the study of this
complexity. The case $d=1$ is studied in \cite{LMMP98}:
it appears that the configuration space is strongly self-similar,
which is certainly also true for $d=2$. Using this self-similarity,
one may hope to obtain some interesting algorithms about the case $d=2$.

\bibliographystyle{alpha}
\bibliography{../Bib/bib}

\end{document}